\documentclass[aps,pra,preprint,superscriptaddress,longbibliography]{revtex4-1}

\usepackage{amsmath,amsthm,amssymb}
\usepackage{xcolor}
\usepackage{graphicx}
\usepackage{siunitx}

\renewcommand{\vec}[1]{\boldsymbol{#1}}
\renewcommand{\tensor}[1]{\tilde{\boldsymbol{#1}}}

\newcommand{\vnabla}{\vec{\nabla}}
\newcommand{\dx}{\;\text{d}\vec{x}}

\def\Jhat{\hat{J}}
\def\op{\hat{\boldsymbol{D}}}

\newcommand{\diff}[2]{\frac{\text{d}#1}{\text{d}#2}}
\newcommand{\idiff}[2]{\text{d}#1 / \text{d}#2}
\newcommand{\pdiff}[2]{\frac{\partial #1}{\partial #2}}
\newcommand{\ipdiff}[2]{\partial #1 / \partial #2}

\begin{document}

\title{
  A fast finite-difference algorithm for topology optimization of permanent magnets
}

\author{Claas Abert}
\email[]{claas.abert@univie.ac.at}
\affiliation{Christian Doppler Laboratory of Advanced Magnetic Sensing and Materials, Faculty of Physics, University of Vienna, Austria}

\author{Christian Huber}
\affiliation{Christian Doppler Laboratory of Advanced Magnetic Sensing and Materials, Faculty of Physics, University of Vienna, Austria}

\author{Florian Bruckner}
\affiliation{Christian Doppler Laboratory of Advanced Magnetic Sensing and Materials, Faculty of Physics, University of Vienna, Austria}

\author{Christoph Vogler}
\affiliation{Faculty of Physics, University of Vienna, Austria}

\author{Gregor Wautischer}
\affiliation{Christian Doppler Laboratory of Advanced Magnetic Sensing and Materials, Faculty of Physics, University of Vienna, Austria}

\author{Dieter Suess}
\affiliation{Christian Doppler Laboratory of Advanced Magnetic Sensing and Materials, Faculty of Physics, University of Vienna, Austria}

\date{\today}

\begin{abstract}
  We present a finite-difference method for the topology optimization of permanent magnets that is based on the FFT accelerated computation of the stray-field.
  The presented method employs the density approach for topology optimization and uses an adjoint method for the gradient computation.
  Comparsion to various state-of-the-art finite-element implementations shows a superior performance and accuracy.
  Moreover, the presented method is very flexible and easy to implement due to various preexisting FFT stray-field implementations that can be used.
\end{abstract}


\maketitle
  
\section{Introduction}
Permanent magnets are a key technology for many industrial applications ranging from sensors \cite{mag_sensor,perm_mag_app} to electric generators and motors \cite{gutfleisch2011magnetic}.
The field generated by these magnets is usually required to have certain properties such as high values and high/low gradients in certain regions.
These properties can be controlled by either designing the magnetization configuration of the magnet or its geometry.
Since the production of magnets with complicated inhomogeneous magnetization configurations is rather involved, the optimization of the geometry of a homogeneously magnetized material is often the most promising approach for field optimization.
The production of arbitrarily shaped magnets has become very affordable due to recent developments in the 3D-printing technology \cite{huber20163d,compton2016direct,lasersinter_mag}.

In numerical mathematics there are several approaches to geometry optimization.
In general there are two classes of methods, namely shape optimization and topology optimization.
For shape optimization the geometry is usually parametrized with a relatively low number of degrees of freedom and optimized with respect to these variables.

Topology optimization is much less restrictive.
As the name suggests, the geometry may even change its topology during optimization.
However, this generality usually comes at the price of a large number of degrees of freedom which leads to high computational costs.
Topology optimization has a long history in the magnetic community \cite{topo_trends}.
Most of the previously presented methods employ the finite-element method for the field calculation.
In this work we present a finite-difference method for shape optimization that uses an FFT accelerated convolution for the field computation.
Compared to previously presented approaches the presented method is exceptionally fast and easy to implement.

\section{Methodology}
We employ the density approach for topology optimization \cite{topo_book}.
In this approach, the topology is described by a scalar indicator function
\begin{equation}
  \rho(\vec{x}) = \begin{cases} 0: & \text{no material}\\ 1: & \text{material} \end{cases}.
\end{equation}
For the optimization process we also allow intermediate values $0 \le \rho \le 1$.
With this indicator function the magnetization field $\vec{M}$ can be written as
\begin{equation}
  \vec{M}(\rho) = \rho^p \vec{M}_0
\end{equation}
where $\vec{M}_0$ is the prescribed magnetization that may be either spatially constant or varying depending on the application.
Note, that we introduced the exponent $p$ as suggested in \cite{topo_book} in order to penalize intermediate values of $\rho$.
For the optimization we consider a general objective function of the form
\begin{equation}
  \Jhat(\rho) = J(\vec{H}(\rho), \rho)
  \label{eq:objective_function}
\end{equation}
that should be minimized.
Here, $\vec{H} = - \vnabla u$ is the magnetic stray field generated by the magnetization $\vec{M}(\rho)$ with $u$ being its scalar potential governed by the Poisson equation
\begin{equation}
  F = \Delta u - \vnabla \cdot \vec{M} = 0
  \label{eq:constraint}
\end{equation}
with open boundary conditions
\begin{equation}
  u(\vec{x}) = \mathcal{O}(1/|\vec{x}|) \quad \text{if} \quad |\vec{x}| \rightarrow \infty.
\end{equation}
A minimium of the objective function $\Jhat$ with respect to the indicator function $\rho$ requires the derivative $\idiff{\Jhat}{\rho}$ to vanish.
The computation of the derivative is also desirable from a numerical point of view since it can be used for iterative minimization with gradient based methods.
The derivative of $\Jhat$ can be written as
\begin{equation}
  \diff{\Jhat}{\rho} =
  \pdiff{\Jhat}{\vec{H}} \diff{\vec{H}}{u} \diff{u}{\rho}
  + \pdiff{\Jhat}{\rho}.
  \label{eq:dj_drho}
\end{equation}
For typical choices of the objective function $\Jhat$, the partial derivatives $\ipdiff{\Jhat}{\vec{H}}$ and $\ipdiff{\Jhat}{\rho}$ can be expressed in a closed analytical form.
However, the computation of $\idiff{u}{\rho}$ is nontrivial since the dependence of the scalar potential $u$ on the indicator function $\rho$ is given by the constraint $F$ which is a partial differential equation.
Numerical computation of $\idiff{u}{\rho}$ by finite differences is possible but infeasible since this procedure requires the solution of $F$ for every degree of freedom of $\rho$ individually.
This shortcoming can be overcome by solution of an adjoint equation \cite{hinze2008optimization}.
Consider the derivative of the constraint $F$
\begin{equation}
  \diff{F}{\rho} =
  \pdiff{F}{u} \diff{u}{\rho} + \pdiff{F}{\rho} = 0.
\end{equation}
Solving for $\idiff{u}{\rho}$, inserting into \eqref{eq:dj_drho}, and applying the adjoint approach yields
\begin{align}
  \diff{\Jhat}{\rho} &=
  \lambda^\ast \pdiff{F}{\rho} + \pdiff{\Jhat}{\rho} \\
  \pdiff{F}{u}^\ast \lambda &=
  - \pdiff{\vec{H}}{u}^\ast \pdiff{\Jhat}{\vec{H}}^\ast 
\end{align}
with $\lambda$ being the so-called adjoint variable.
Inserting \eqref{eq:constraint} results in the system
\begin{align}
  \diff{\Jhat}{\rho} &= p \rho^{p-1} \vec{M}_0 \cdot \vnabla \lambda + \pdiff{\Jhat}{\rho} \label{eq:adjoint1}\\
  \Delta \lambda     &= \vnabla \cdot \pdiff{\Jhat}{\vec{H}}. \label{eq:adjoint2}
\end{align}
Note that \eqref{eq:adjoint2} has exactly the same form as the constraint \eqref{eq:constraint}, i.e. the right-hand side of the Poisson equation is given as the divergence of a vector entity.
Moreover, the knowledge of the gradient of the adjoint variable $\vnabla \lambda$ is sufficient for the computation of the derivative \eqref{eq:adjoint1}.
This means that both the forward problem $F$ as well as the adjoint problem \eqref{eq:adjoint1} and \eqref{eq:adjoint2} can be expressed in terms of the stray-field operator
\begin{equation}
  \op: C^0(\mathbb{R}^3, \mathbb{R}^3) \rightarrow 
       C^0(\mathbb{R}^3, \mathbb{R}^3)
\end{equation}
with
\begin{equation}
  \vec{H} = \op(\vec{M})
\end{equation}
that maps the magnetization vector field $\vec{M}$ onto the vector field $\vec{H}$.
With this definition the objective function and its derivative can be written as
\begin{align}
  \Jhat              &= J( \op[\vec{M}(\rho)], \rho ) \\
  \diff{\Jhat}{\rho} &= - p \rho^{p-1} \vec{M}_0 \cdot \op\left( \pdiff{\Jhat}{\vec{H}} \right) + \pdiff{\Jhat}{\rho} 
\end{align}
This formulation can be readily used with arbitrary numerical methods for the stray-field computation to perform topology optimization.

\section{Discretization}
Various numerical algorithms for the discrete computation of the stray-field have been proposed, see e.g. \cite{abert2013numerical}.
Among the fastest and most accurate algorithms is the fast-Fourier-transform (FFT) accelerated convolution with the demagnetization tensor $\tensor{N}$.
The prerequisite for this method is a regular cuboid grid that enables the formulation of the demagnetization-field problem as a discrete convolution
\begin{align}
  \vec{H}_{\vec{i}} &=
  \sum_{\vec{j}} \tensor{N}_{\vec{i} - \vec{j}} \vec{m}_{\vec{j}} \label{eq:convolution}\\
  \tensor{N}_{\vec{i} - \vec{j}} &=
  - \frac{1}{4 \pi V_\text{cell}} \int_{\Omega_{\vec{i}}} \int_{\Omega_{\vec{j}}} \vnabla \vnabla' \frac{1}{|\vec{x} - \vec{x}'|} \dx \dx' \label{eq:demag_tensor}
\end{align}
where $V_\text{cell}$ is the volume of a single simulation cell and $\Omega_{\vec{i}}$ and $\Omega_{\vec{j}}$ are the simulation cells at multiindex $\vec{i}$ and $\vec{j}$ respectively.
Note that, due to the regularity of the grid, the sixfold integral only depends on the difference of multiindices $\vec{i}$ and $\vec{j}$ and not on their specific values.
While a naive implementation of the convolution \eqref{eq:convolution} would require a computational complexity of $\mathcal{O}(N^2)$, the computation in Fourier space and application of the FFT reduces the complexity to $\mathcal{O}(N \log N)$.
This procedure, including the accurate computation of the discrete demagnetization tensor \eqref{eq:demag_tensor} and the optimal implementation of the fast convolution is well documented, e.g. in micromagnetic literature \cite{yuan1992fast,hayashi1996calculation,donahue2009parallelizing}.
Moreover, various open-source implementations exist that can be used to implement the presented topology-optimization strategy \cite{donahue2009parallelizing,vansteenkiste2014design,abert2015full}.
For this work, we use the CPU code of the micromagnetic simulator magnum.fd \cite{magnumfd}.
Minimization of the objective function is performed with a quasi-Newton method that is able to handle the constraint $0 \le \rho \le 1$ of the indicator function.

In order to compare our implementation with respect to accuracy and performance, we implement two additional methods based on the finite-element method (FEM).
Finite-element methods solve the Poisson equation \eqref{eq:constraint} by means of a variational approach and work on arbitrary tetrahedral meshes.
However, the treatment of the required open boundary conditions is nontrivial with FEM.
For the first FEM approach we extend the mesh beyond the region of interest that is used for the topology optimization.
The size of the extended mesh is chosen to be approximately 5 times larger than the original mesh in each spatial dimension and we apply zero Dirichlet boundary conditions on the outer boundary.
This so-called truncation approach was already shown to provide good results for topology optimization in \cite{huber2017topology}.
We solve the stray-field potential $u$ by the weak formulation
\begin{equation}
  \int_{\Omega_\text{all}} \vnabla u \cdot \vnabla v \dx =
  \int_{\Omega_\text{mag}} \vec{M} \cdot \vnabla v \dx
  \label{eq:fem_weak}
\end{equation}
where $\Omega_\text{all}$ is the complete meshed region, $\Omega_\text{mag}$ is the magnetic region and the trial and test functions $u$ and $v$ are discretized with piecewise affine, globally continuous functions $u, v \in \mathcal{P}^1$.
Since the stray-field operator $\op$ is used for both the forward problem and the adjoint problem, the discrete version of $\op$ should use the same function space for the output $\vec{H}$ as for the input $\vec{M}$.
The stray-field $\vec{H}$ is given by the negative gradient of the scalar potential $u$.
With $u$ being a piecewise affine function, the field $\vec{H}$ is naturally given as a componentwise piecewise constant, globally discontinuous function $H_i \in \mathcal{P}^0$.
As \eqref{eq:fem_weak} does not pose any requirements on the differentiability of the magnetization $\vec{M}$, we choose both the input and the ouput function of $\op$ to be componentwise $\mathcal{P}^0$.

The downside of the truncation approach is the requirement of additional mesh nodes which increases both the storage requirements as well as the computational costs.
The additional mesh nodes can be avoided by application of a hybrid finite-element/boundary-element method (FEM/BEM) \cite{fredkin1990hybrid}.
We use the same function spaces as for the pure FEM truncation approach.
For the BEM part we use different implementations, namely a collocation approach \cite{lindholm1984three} and a Galerkin approach with and without matrix compression via $\mathcal{H}$-matrices \cite{hackbusch2015hierarchical}.

For the FEM implementation we use the multipurpose library FEniCS \cite{alnaes2015fenics}, for the BEM implementation we use BEM++ \cite{smigaj2015solving} and for $\mathcal{H}$-matrix compression H2Lib \cite{h2lib}.
The minimization for all methods is done with the L-BFGS-B minimizer of the SciPy library \cite{scipy}.
In the following we will refer to the truncation approach as FEM and to the hybrid method as FEM/BEM.
The FFT accelerated method will be referred to as FD.

\section{Validation and Benchmarks}
\begin{figure}
  \includegraphics{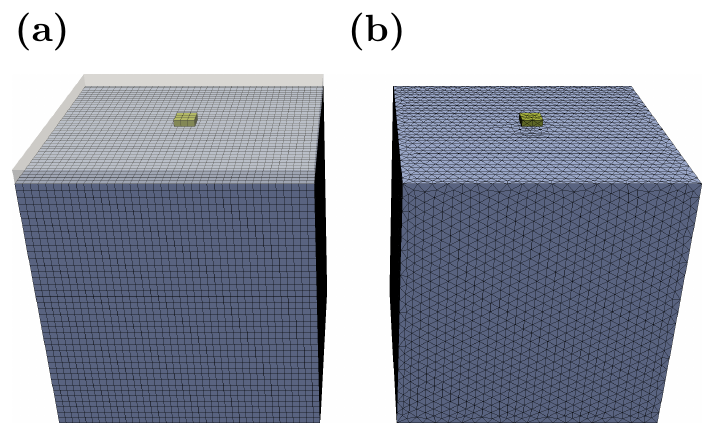}
  \caption{
    Geometry for the topology-optimization benchmark problem.
    The large blue cube marks the region considered for topology optimization.
    The $z$-component of the stray field is maximized in the green box.
    (a) Finite-difference mesh with \num{62361} cells.
    (b) Finite-element mesh with \num{13072} nodes and \num{60189} cells.
  }
  \label{fig:mesh}
\end{figure}
For validation and benchmarking purposes, we consider a simple test problem.
We aim to maximize the $z$-component of the stray field in a small box above a unit cube with magnetization $\vec{M}_0 = (0,0,1)$ that is considered for topology optimization, see Fig.~\ref{fig:mesh}.
For the FEM method we add an external mesh which increases the number of mesh nodes from \num{13072} to \num{25055} and the number of cells from \num{60189} to \num{146522}.
The objective function for the problem reads
\begin{equation}
  \Jhat = - \frac{1}{2} \int_{\Omega} H_z^2 \dx
\end{equation}
where $\Omega$ is the region where the field is maximized.
The corresponding derivative reads
\begin{equation}
  \diff{\Jhat}{\rho} = p \rho^{p-1} \vec{M}_0 \cdot \op\left( \chi_\Omega H_z \right)
\end{equation}
with $\chi_\Omega$ being the characteristic function of the region $\Omega$.
We set $p = 3$ and choose $\rho(\vec{x}) = 1$ as start condition for the iterative optimization.

\begin{figure}
  \includegraphics{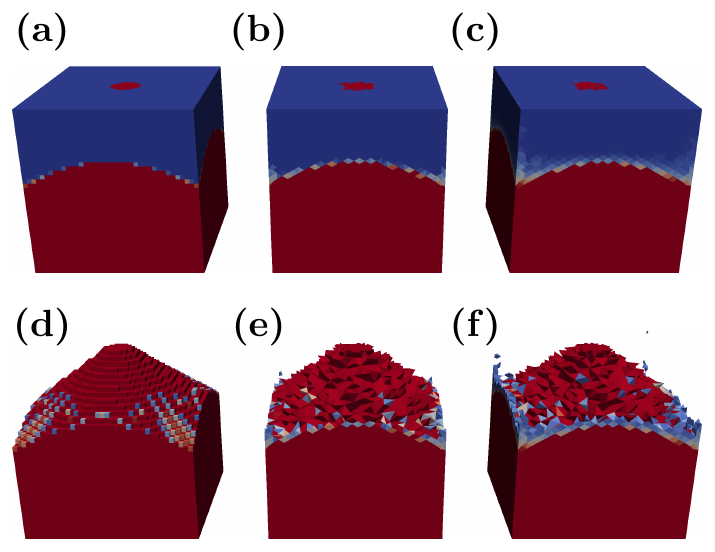}
  \caption{
    Optimized topology for maximum $z$-component of the stray-field in a small box above the optimization region that is magnetized in $z$-direction.
    The results for the presented finite-difference algorithm (a),(d) is shown along with the results for FEM (b),(e) and collocation FEM/BEM (c),(d).
    (a)--(c) Indicator function $\rho$ in the optimization region (red = 1, blue = 0).
    (d)--(f) Optimized geometry ($\rho > 0.1$).
  }
  \label{fig:rho}
\end{figure}
Figure~\ref{fig:rho} shows the resulting topology as computed with the presented finite-difference algorithm compared to the results computed with FEM and collocation FEM/BEM.
For all approaches the same L-BFGS-B method with identical tolerances was applied.
At a first glance the quality of the FD solution seems better than those of the FEM and FEM/BEM solutions.
The regular cuboid grid leads to a relatively smooth representation of the optimized geometry compared to the other methods.
Furthermore, there seem to be less simulation cells with intermediate values for the indicator function $\rho$.
\begin{figure}
  \includegraphics{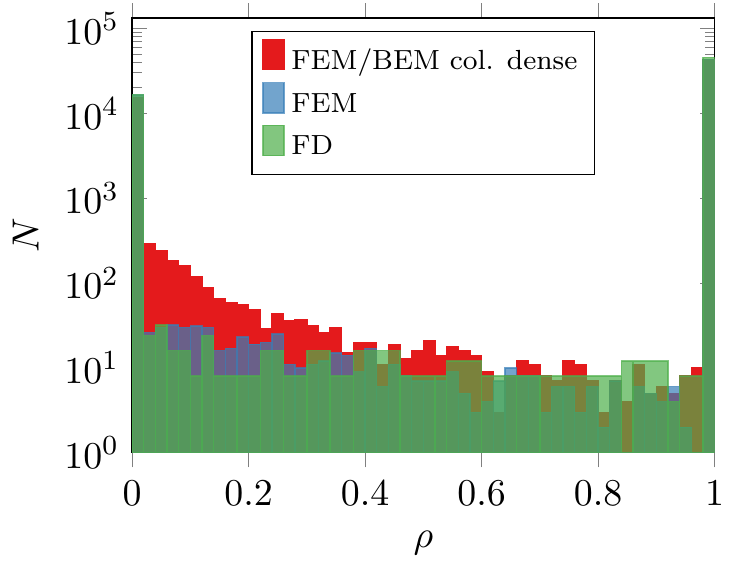}
  \caption{
    Cell distribution of indicator function for the optimized topology computed with various methods.
    The number of cells $N$ is plotted against the value of $\rho$.
  }
  \label{fig:histogram}
\end{figure}
This impression is confirmed by Fig.~\ref{fig:histogram} that shows the cell distribution of the indicator function $\rho$ for the different methods.
For this particular problem, every individual simulation cell either increases or decreases the objective function.
Hence, intermediate values of $\rho$ indicate an inaccurate simulation result.
A possible reason for this behaviour is the accuracy of the discrete stray-field operator $\op$.
If the operator is inaccurate it might not reflect the self-adjoint properties of the original problem.
Thus, the gradient computed by \eqref{eq:adjoint2} might not accurately fit the objective function \eqref{eq:objective_function} which leads to bad convergence of the iterative optimization procedure.

\begin{figure}
  \includegraphics{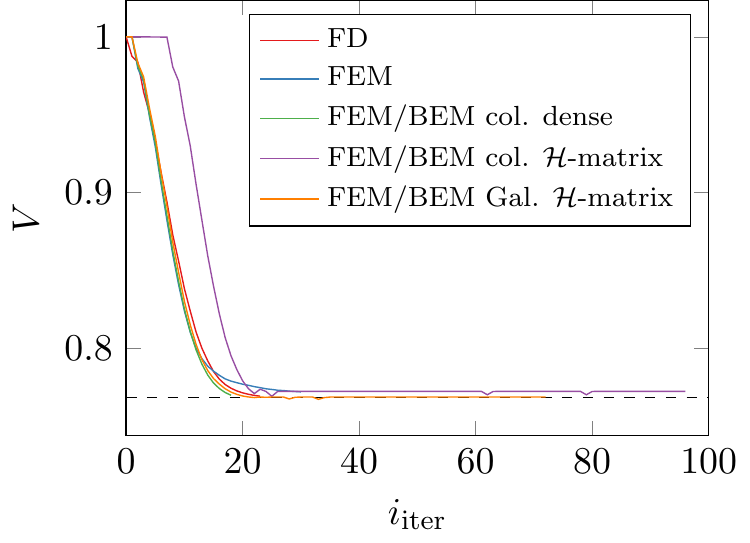}
  \caption{
    Convergence speed comparison of the topology optimization with various stray-field methods.
    The volume fraction of the optimized volume $V$ is plotted against the number of right-hand-side evaluations $i_\text{iter}$.
    The black dashed line denotes the reference solution computed with the FD method and a 4 times higher resolution in every spatial dimension.
  }
  \label{fig:convergence}
\end{figure}
Figure~\ref{fig:convergence} shows a comparison of the convergence for the individual methods.
In the beginning, all methods except the collocation FEM/BEM with $\mathcal{H}$-matrix compression show almost the same descent velocity in the volume fraction of the optimized topology.
Both FEM/BEM methods with $\mathcal{H}$-matrix compression require significantly more function evalutions to converge than the remaining methods.
Obviously, the gradient computation is very sensitive to approximations in the stray-field computation.
The FEM method converges only slightly slower than FD and collocation FEM/BEM with dense BEM matrix.
However, it saturates at a significantly higher volume fraction than the reference solution.

\begin{table}
  \centering
  \begin{tabular}{
      l
      S[table-format=2.0]
      S[table-format=2.1]
      S[table-format=1.3]
      S[table-format=2.1]
    }
    \toprule
    \multicolumn{1}{c}{Method}
    & \multicolumn{1}{c}{$N_\text{iter}$}
    & \multicolumn{1}{c}{$T_\text{stray} [\si{ms}]$}
    & \multicolumn{1}{c}{$T_\text{iter}  [\si{s}]$}
    & \multicolumn{1}{c}{$T_\text{total} [\si{s}]$} \\
    \colrule
    FD                                & 25 & 31.0 & 0.196 &  4.9 \\
    FEM                               & 31 & 57.0 & 2.303 & 71.4 \\
    FEM/BEM col. dense                & 19 & 69.0 & 0.826 & 15.7 \\
    FEM/BEM col. $\mathcal{H}$-matrix & 97 & 49.6 & 0.684 & 66.4 \\
    FEM/BEM Gal. $\mathcal{H}$-matrix & 73 & 62.1 & 0.781 & 57.0 \\
    \botrule
  \end{tabular}
  \caption{
    Comparison of convergence and timings for various stray-field methods.
    $N_\text{iter}$ denotes the total number of L-BFGS-B iterations, $T_\text{stray}$ denotes the time for a single stray-field computation, $T_\text{iter}$ denotes the average time of a single L-BFGS-B iteration, and $T_\text{total}$ denotes the total time for the optimization excluding setup time.
  }
  \label{tab:comparison}
\end{table}
Table~\ref{tab:comparison} shows the timings for the individual methods.
All simulations were carried out as single threaded on a standard laptop computer with an Intel Core i7 \SI{2.90}{GHz} CPU and \SI{8}{GB} RAM.
The finite-element matrices were solved with the sparse direct solver MUMPS \cite{amestoy2000mumps}.
The setup time required for the assembly of the demagnetization tensor and the FEM matrices as well as the matrix factorization for the direct solver were excluded from the timings.
By far the fastest method is the presented FD method that outperforms all other methods both in terms of stray-field computation as well as total time.
The second fastest method is the collocation FEM/BEM with dense BEM matrix.
However, we recall that the optimization result of this method is of bad quality as it contains a large number of intermediate values for $\rho$.
All other methods are at least a factor of 10 slower than the presented method.
\begin{figure}
  \includegraphics{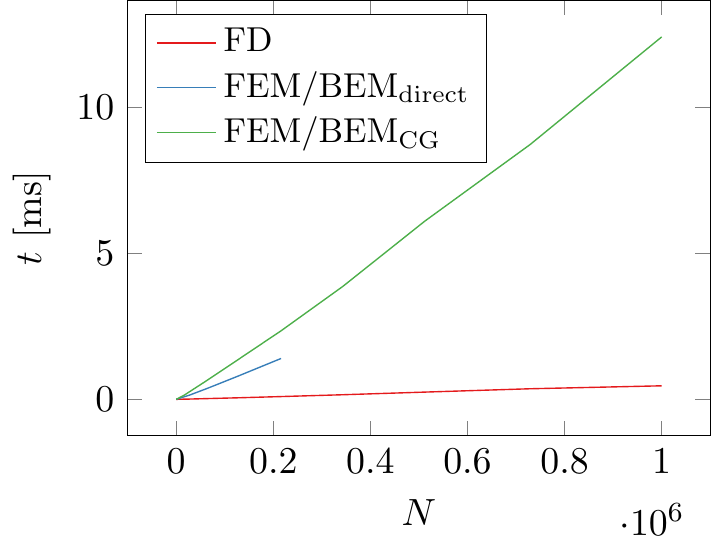}
  \caption{
    Scaling of the stray-field computation time of the FD method compared to FEM/BEM with direct solver and conjugate gradient solver respectively.
    The computation time $t$ is plotted agains the numbers of degrees of freedom $N$ (cells in the case of FD and nodes in the case of FEM/BEM).
  }
  \label{fig:demag_scaling}
\end{figure}
The performance gain of the FD method compared to the FEM/BEM method will even be more significant for larger systems as suggested by Fig.~\ref{fig:demag_scaling}.
Both methods seem to scale approximately linear which can be explained by the algorithmic complexity of both methods of $\mathcal{O}(N \log N)$.
However, at large problem sizes the FD method offers a tremendous performance gain compared to FEM/BEM.
For \num{e6} degrees of freedom, FD is more than a factor of 25 faster than FEM/BEM.
Note, that the timings for FEM/BEM with a direct solver could only by determined for small systems, since the memory consumption of the matrix factorization exceeded the capabilities of the test machine for larger systems.

Note also that Fig.~\ref{fig:demag_scaling} shows the computation time of FEM/BEM with respect to the number of mesh nodes instead of cells, as the finite-element systems for the potential $u$ scales with the number of nodes.
However, even if instead the number of cells is considered for the FEM/BEM scaling, the FD method outperforms FEM/BEM by a factor of 5, while still being more accurate.

\section{Optimization of field and gradient}
\begin{figure}
  \includegraphics{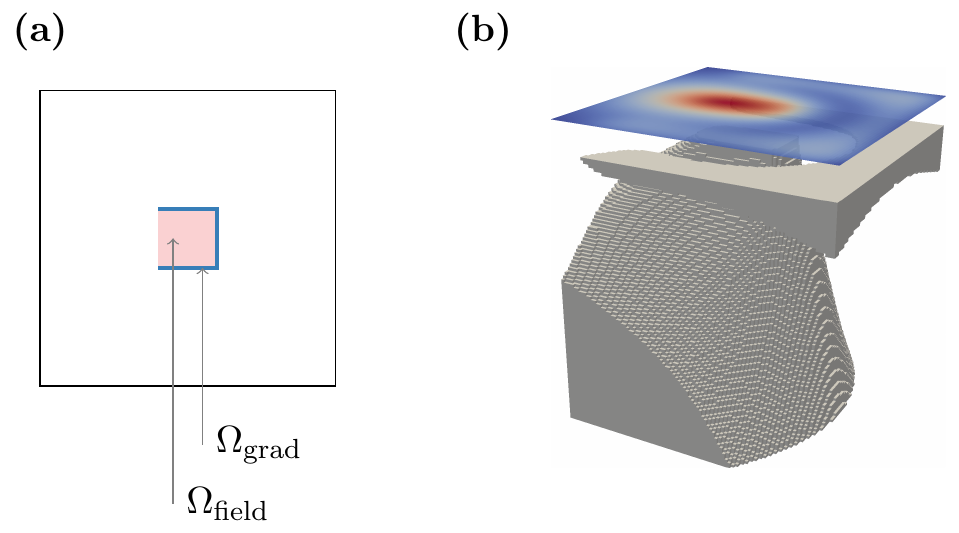}
  \caption{
    Maximization of field and gradient on a plane above the magnet.
    (a) Top-view of the areas of field and gradient optimization.
    (b) Optimized geometry and $z$-component of the resulting stray-field (red = large values, blue = small values)
  }
  \label{fig:head}
\end{figure}
In a more complex example the field and its gradient should be optimized in certain areas of a plane above the region subject to topology optimization, see Fig.~\ref{fig:head}\,(a).
The objective function for this experiment reads
\begin{equation}
  \Jhat = - \frac{1}{2} \int_{\Omega_\text{field}} H_z^2 \dx
          - \frac{\beta}{2} \int_{\Omega_\text{grad}} \left( \diff{H_z}{\vec{n}} \right)^2 \dx
\end{equation}
where $\vec{n}$ is the outward-pointing unit vector of the region of maximum field and $\beta$ is chosen as \num{e5}.
The region of maximum field $\Omega_\text{field}$ has a size of \num{20x20} simulation cells and the region of maximum gradient $\Omega_\text{grad}$ has a width of 1 simulation cell and marks three sides of the maximum field region, see Fig.~\ref{fig:head}\,(a).
The resulting derivative of the objective function reads
\begin{equation}
  \diff{\Jhat}{\rho} = p \rho^{p-1} \vec{M}_0 \cdot \op\left(
      \chi_{\Omega_\text{field}} H_z
    + \beta \chi_{\Omega_\text{grad}} \frac{\text{d}^2 H_z}{\text{d} \vec{n}^2}
  \right)
\end{equation}
with $\chi_{\Omega_\text{field}}$ and $\chi_{\Omega_\text{grad}}$ being the characteristic functions of $\Omega_\text{field}$ and $\Omega_\text{grad}$ respectively.
We use a finite-difference three-point stencil for the approximation of the second derivative of $H_z$.
Similar to the simple test problem in the preceding section we choose $\vec{M}_0 = (0,0,1)$ and $p = 3$ and perform the optimization with a L-BFGS-B method.
The optimization result is depicted in Fig.~\ref{fig:head}\,(b).

\section{Conclusion}
We present a fast and accurate finite-difference method for topology optimization of permanent magnets with respect to their stray field.
The implementation of the method is simple due to the possibility to use existing highly optimized codes for the computation of the magnetic stray field.
We compare the method to various finite-element implementations and demonstrate that the presented method is significantly faster and more accurate than any finite-element implementation.
For typical applications the possibly irregular mesh used by the finite-element method is considered an advantage over the regular cuboid grid that is required by the finite-difference method.
However, this advantage only exists for predefined geometries where irregular meshes are able to accurately approximate complex structures with a relatively low number of nodes.
For topology optimization the geometry is not known upfront and thus a regular mesh might even be favorable because of the simple geometric representation.
The presented method is general and can easily be extended by additional terms to the objective function such a volume constraints or higher order derivatives.

\section*{Acknowledgements}
The financial support by
the Austrian Federal Ministry of Science, Research and Economy and the National Foundation for Research, Technology and Development
as well as
the Austrian Science Fund (FWF) under grant F4112 SFB ViCoM,
and the Vienna Science and Technology Fund (WWTF) under grant MA14-44,
is gratefully acknowledged.

\end{document}